\documentclass[11pt,english,a4paper,floatfix]{article}
\usepackage{jcappub}

\pdfoutput=1
\usepackage{bm}
\usepackage{amsfonts}
\usepackage{subfigure}

\newcommand{\be}{\begin{equation}}
\newcommand{\ee}{\end{equation}}
\newcommand{\bea}{\begin{eqnarray}}
\newcommand{\eea}{\end{eqnarray}}

\renewcommand{\vec}[1]{\boldsymbol{#1}}

\newcommand{\adotovera}{\frac{\dot a}{a}}

\usepackage{lmodern}
\usepackage{slantsc}
\newcommand{\CONCEPT}{\textsc{co\textsl{n}cept}}
\newcommand{\CLASS}{\textsc{class}}
\newcommand{\CAMB}{\textsc{camb}}

\newcommand{\PKDGRAV}{\textsc{pkdgrav}{\footnotesize 3}}

\begin{document}


\title{Dark energy perturbations in $N$-body simulations}

\author[a]{Jeppe Dakin,}
\author[a]{Steen Hannestad,}
\author[b]{Thomas Tram,}
\author[c]{Mischa Knabenhans,}
\author[c]{Joachim Stadel}

\affiliation[a]{Department of Physics and Astronomy, Aarhus University,
 DK--8000 Aarhus C, Denmark}
\affiliation[b]{Aarhus Institute of Advanced Studies (AIAS), Aarhus University, DK--8000 Aarhus C, Denmark}
\affiliation[c]{Institute for Computational Science, University of Zurich, CH--8057 Z{\"u}rich, Switzerland}

\emailAdd{dakin@phys.au.dk}
\emailAdd{sth@phys.au.dk}
\emailAdd{thomas.tram@aias.au.dk}
\emailAdd{mischak@physik.uzh.ch}
\emailAdd{stadel@physik.uzh.ch}

\abstract{
We present $N$-body simulations which are fully compatible with general relativity, with dark energy consistently included at both the background and perturbation level. We test our approach for dark energy parameterised as both a fluid, and using the parameterised post-Friedmann (PPF) formalism. 
In most cases, dark energy is very smooth relative to dark matter so that its leading effect on structure formation is the change to the background expansion rate. This can be easily incorporated into Newtonian $N$-body simulations by changing the Friedmann equation. 
However, dark energy perturbations and relativistic corrections can lead to differences relative to Newtonian $N$-body simulations at the tens of percent level for scales $k < (10^{-3}$ -- $10^{-2})\,\text{Mpc}^{-1}$, and given the accuracy of upcoming large scale structure surveys such effects must be included.
In this paper we will study both effects in detail and highlight the conditions under which they are important.
We also show that our $N$-body simulations exactly reproduce the results of the Boltzmann solver \CLASS{} for all scales which remain linear. 
}

\maketitle


\section{Introduction}

In the coming few years, new, large galaxy surveys such as those from LSST \cite{LSST} and EUCLID \cite{EUCLID} will provide extremely precise measurements of the large scale structure of our Universe.
While such surveys will allow probing of e.g.\ the dark energy component and the mass of neutrinos with unprecedented precision they also put stringent requirements on numerical simulations of large scale structure
(see e.g.\ \cite{Schneider:2015yka} for a detailed discussion of this subject).

Among the effects which need to be included are massive neutrinos and photons, as well as effects from general relativity.
In a series of papers we have discussed and developed a framework for incorporating this into $N$-body simulations
(see \cite{Tram:2018znz,Fidler:2015npa,Fidler:2016tir,Brandbyge:2016raj,Adamek:2017grt}\textbf{}).
This involves treating massive (but light) neutrinos, as well as photons and effects on the metric in linear perturbation theory using the \CLASS{} \cite{Blas:2011rf} code and subsequently add them as source terms in the dark matter equations of motion in the simulation.

Given that most models of dark energy predict that it is at most moderately inhomogeneous on scales below the horizon we can utilize the same approach for this component, and the purpose of this paper is indeed to demonstrate that our approach works extremely well for two standard parameterisations of dark energy; dark energy as a fluid and dark energy in the parameterised post-Friedmann approach\footnote{The approach should generalize to other types of dark energy which obey the conditions that: 1) Dark energy couples only gravitationally to other species, 2) dark energy perturbations remain linear at all times so that they can be adequately treated using linear perturbation theory.} \cite{Hu:2004kh,Fang:2008sn}.

In models where dark energy only has gravitational coupling to the matter fields, its effect on structure formation can basically be separated into two components: 1) The presence of dark energy changes the expansion of the background, leading to different growth rates of fluctuations in other species.
2) Unless dark energy is in the form of a cosmological constant, $\Lambda$, it contains perturbations, and these act as a source term for gravitational clustering.

Given that dark energy is very smooth relative to dark matter, in most cases the leading effect is the change to the background expansion rate. This can be easily incorporated into Newtonian $N$-body simulations by simply changing the Friedmann equation. However, for simulations with percent-level accuracy the second effect must also be taken into account, in particular in models where the effective dark energy sound speed is small.
In this paper we will study both effects in detail and highlight the conditions under which they are important using the numerical framework presented in \cite{Tram:2018znz}.

The paper is structured as follows: In Section 2 we discuss the theoretical set-up needed to include dark energy in both the fluid and PPF approaches. In Section 3 we present our numerical results, and finally Section 4 contains a discussion and our conclusions.

\section{Method and implementation}

Our treatment closely follows that of \cite{Tram:2018znz} (see also \cite{Fidler:2015npa,Fidler:2016tir,Brandbyge:2016raj,Adamek:2017grt}), but for clarity we briefly reiterate the needed steps here.

For pure matter (i.e.\ a pressureless component) the continuity and Euler equations for the density contrast $\delta_{\text{m}}$ and peculiar velocity $\vec{v}_{\text{m}}$ can be written as 
\begin{align}
\dot\delta_{\text{m}} ^\text{Nb}+ \nabla \cdot {\vec v}_{\text{m}}^\text{Nb} &= 0\,, \\
( \partial_\tau + {\cal H} ) {\vec v}_{\text{m}}^\text{Nb} &= -\nabla \phi + \nabla \gamma^\text{Nb}\,,
\end{align}
where a dot denotes differentiation with respect to conformal time $\tau$ and ${\cal H} = \dot{a}/a$ is the conformal Hubble parameter with $a$ being the cosmic scale factor. The superscript `Nb' denotes quantities in the $N$-body gauge. The quantity $\gamma^\text{Nb}$ is a correction to the Euler equation that originates in perturbed non-dust components such as relativistic species and dark energy. The gauge-invariant potential $\phi$ satisfies a Newtonian Poisson equation in the $N$-body gauge, but with contributions from all species, i.e.\
\begin{equation}
	\nabla^2\phi =4\pi G a^2 \sum_\alpha \delta\rho_\alpha^\text{Nb}\,,
\end{equation}
with $\alpha\in\{\text{cdm}, \text{b}, \gamma, \nu, \text{DE}\}$ running over all species.

From \cite{Fidler:2017pnb}, the Fourier space equation for $\gamma^\text{Nb}$ can be written as
\begin{equation}
k^2 \gamma^\text{Nb}  = -(\partial_\tau + {\cal H} ) \dot H_\text{T} ^\text{Nb} + 8 \pi G a^2 \Sigma\,, \label{eq:gamma_from_perturbations}
\end{equation}
where $\Sigma$ is the total anisotropic stress of all species and $H_\text{T}^\text{Nb}$ is the trace-free component of the spatial part of the metric in $N$-body gauge
(see e.g.\ \cite{Adamek:2017grt}).
For species other than dark energy the calculation of $\gamma^\text{Nb}$ can be found outlined in e.g.\ the appendix of \cite{Tram:2018znz}.
For the dark energy component (as for the other components) the quantities we shall need are $\delta \rho$, $\delta p$ and $\sigma$.
In the next subsection we will discuss how to extract these quantities in two different formulations of dark energy; dark energy as a fluid and dark energy in the parameterised post-Friedmann (PPF) approach.

Following the same steps as in \cite{Tram:2018znz} we split the total ``force-potential'' $\phi - \gamma^\text{Nb}$ experienced by particles in the actual simulation into a contribution coming from the matter itself (calculable using standard techniques in the $N$-body simulation), $\phi_{\text{sim}}$, and  contributions coming from other species (neutrinos, photons and dark energy) and the GR correction $\gamma^\text{Nb}$, $\phi_{\text{GR}}$:
\begin{equation}
	\phi - \gamma^\text{Nb} \equiv \phi_{\text{sim}} + \phi_{\text{GR}}\,,
\end{equation}
with $\phi_{\text{GR}}$ given by
\begin{align}
	\nabla^2\phi_{\text{GR}} &\equiv \nabla^2\bigl( \phi_\gamma + \phi_\nu + \phi_{\mathrm{DE}} - \gamma ^\text{Nb}\bigr) \notag \\
	&\equiv 4 \pi G a^2 \bigl( \delta\rho_{\gamma}^\text{Nb} + \delta\rho_{\nu}^\text{Nb} + \delta\rho_{\mathrm{DE}}^\text{Nb} + \delta\rho_{\text{metric}} \bigr) \label{eq:Poissson_GR} \\
	&\equiv 4\pi G a^2 \delta\rho_{\text{GR}}^\text{Nb}\,. \notag
\end{align}
Here $\delta\rho_{\text{metric}}$ is a fictitious density perturbation which amounts to the GR potential correction $\gamma^\text{Nb}$;
\begin{equation}
	\nabla^2 \gamma^\text{Nb} = -4\pi Ga^2 \delta\rho_{\text{metric}}\,. \label{eq:gamma_Poisson}
\end{equation}
Following the same prescription as in \cite{Brandbyge:2016raj},
at each time step in the simulation we realise\footnote{Note that we assume adiabatic initial conditions so that the set of random numbers used for the realisation is the same for all species, including the metric term $\gamma$.} $\delta\rho_{\text{GR}}^{\text{Nb}}$ in Fourier space, solve its Poisson equation \eqref{eq:Poissson_GR}, transform to real space and apply the force from $\phi_{\text{GR}}$ to the matter particles, in addition to the usual force from the matter particles themselves (corresponding to $\phi_\text{sim}$).

As described in \cite{Tram:2018znz}, to compute $\delta\rho_{\text{GR}}$ in linear perturbation theory, a \CLASS{} computation has been run in advance, providing us with $\delta \rho_\gamma$, $\delta \rho_\nu$ and $\delta \rho_{\rm DE}$ in either synchronous or conformal Newtonian gauge, as well as $\dot{H}_{\text{T}}^\text{Nb}$ and $\Sigma$, all as functions of $a$ and $k$. From these we can calculate $\delta \rho$ for all species (including the metric) in $N$-body gauge. All of these are subsequently realised on a grid in real space using the formalism outlined in \cite{Dakin:2017idt}.

In conclusion, all that is needed in order to run simulations with dark energy perturbations consistent with GR is to calculate $\delta \rho_{\rm DE}^{\text{Nb}}(a,k)$, $\delta p_{\rm DE}^{\text{Nb}} (a,k)$, and $\sigma_{\rm DE} ^{\text{Nb}}(a,k)$. These depend crucially on the way in which the dark energy component is parameterised, as we will now discuss.

\subsection{Dark energy parameterisation}

While the true nature of the dark energy component responsible for the current acceleration of the expansion is unknown, a vast number of models for it exist.
Possible realisations of dark energy can be in the form of scalar fields evolving in very flat potentials, i.e.\ quintessence-like models. 

From an ``effective theory'' point of view, two parameterisations are particularly popular, and have been implemented in both \CLASS{} \cite{Blas:2011rf} and \CAMB{} \cite{Lewis:1999bs}: Dark energy as a fluid, and dark energy in the parameterised post-Friedmann approach. These two models are representative of a wide range of different physical models, including quintessence and many modified gravity models. 
We therefore restrict our treatment to these two models. However, as noted before, any dark energy model in which the dark energy couples only gravitationally to other fields, and where the dark energy remains linear at all times and on all scales should be treatable using our prescription.

\subsubsection{Dark energy as a fluid}

A simple and often used parameterisation of dark energy is to describe it as a fluid with equation of state $w(a)$ and constant rest-frame sound speed $c_{\mathrm{s}}$ (see e.g.\ \cite{Ballesteros:2010ks} for a detailed description). In this paper we make use of the parameterisation $w(a) = w_0 + w_a(1 - a)$, though none of the equations presented rely on this choice.

In synchronous gauge (superscript `s') the continuity and Euler equations of a dark energy fluid take the form
\begin{align}
\dot\delta_{\text{DE}}^{\text{s}} &= -(1+w)\biggl(\theta_{\text{DE}}^{\text{s}}+\frac{\dot h}{2}\biggr) - 3 (c_{\mathrm{s}}^2-w) {\cal H} \delta^{\text{s}}_{\text{DE}} - 9 (1+w)(c_{\mathrm{s}}^2-c_{\mathrm{a}}^2) {\cal H}^2 \frac{\theta^{\text{s}}_{\text{DE}}}{k^2} \,, \\
\dot \theta^{\text{s}}_{\text{DE}} &= -(1-3 c_{\mathrm{s}}^2) {\cal H} \theta^{\text{s}}_{\text{DE}} + \frac{c_{\mathrm{s}}^2 k^2}{1+w} \delta^{\text{s}}_{\mathrm{DE}} - k^2 \sigma_{\text{DE}} \,, \label{eq:fluid_euler}
\end{align}
where $c_{\mathrm{a}}^2 \equiv \dot p_{\text{DE}}/\dot \rho_{\text{DE}}$ is the adiabatic sound speed squared. The standard assumption is to take $\sigma_{\text{DE}}=0$, i.e.\ to have no anisotropic stress for the dark energy. These equations are implemented in standard versions of codes like \CAMB{} and \CLASS{}, and adequately describe a range of dark energy models.

We shall need the pressure perturbation from the dark energy fluid in order to calculate its contribution to $\gamma$. This is given by
\begin{equation}
\frac{\delta p_{\text{DE}}}{\rho_{\text{DE}}} = c_{\mathrm{s}}^2 \delta_{\text{DE}} + 3 {\cal H} (1+w) (c_{\mathrm{s}}^2-c_{\mathrm{a}}^2) \frac{\theta_{\text{DE}}}{k^2} \,.
\label{eq:fluidp}
\end{equation}
As seen from \eqref{eq:fluid_euler}, models in which there are phantom crossings (i.e.\ $1+w$ crosses 0) become pathological in this description. In that case a simple extension of the fluid concept is to use what is commonly referred to as the parameterised post-Friedmann approach.

\subsubsection{Parameterised Post-Friedmann dark energy}

The Parameterised Post-Friedmann (PPF) description of dark energy~\cite{Hu:2008zd,Fang:2008sn} is the standard implementation of phantom crossing dark energy models in both \CLASS{} and \CAMB{}. We will now sketch the derivation of the PPF formalism, but we refer the reader to the references above for a more complete derivation. We start by defining the dark energy (DE) perturbation in the DE rest frame,
\begin{equation}
k^2 \Gamma \equiv -4 \pi G a^2  \delta \rho_{\mathrm{DE}}^{\rm rest} \, ,
\end{equation}
where we have assumed spatial flatness for simplicity (the generalisation to non-flat space is straightforward, however). We now consider the Poisson equation in conformal Newtonian gauge (superscript `N'),
\begin{align}
k^2 \phi &= -4 \pi G a^2 \biggl(\delta\rho^{\mathrm{N}}_{\text{tot}} -3 \mathcal{H} (\rho_{\text{tot}} + p_{\text{tot}}) \frac{\theta^{\mathrm{N}}_{\text{tot}}}{k^2} \biggr) \\
&= -4 \pi G a^2 \biggl(\delta \rho^{\mathrm{N}}_{\mathrm{t}} -3 \mathcal{H} (\rho_{\mathrm{t}} + p_{\mathrm{t}}) \frac{\theta_{\mathrm{t}}^{\mathrm{N}}}{k^2} + \delta \rho^{\mathrm{N}}_{\mathrm{DE}} -3 \mathcal{H} (\rho_{\mathrm{DE}} + p_{\mathrm{DE}}) \frac{\theta_{\mathrm{DE}}^{\mathrm{N}}}{k^2} \biggr) \\
&=k^2 \Gamma  -4 \pi G a^2 \biggl(\delta \rho^{\mathrm{N}}_{\mathrm{t}} -3 \mathcal{H} (\rho_{\mathrm{t}} + p_{\mathrm{t}}) \frac{\theta_{\mathrm{t}}^{\mathrm{N}}}{k^2} \biggr)\, ,
\end{align}
where we have separated the DE contribution from all other species (subscript `t'). The requirement that we recover the usual Newtonian Poisson equation in the small-scale limit forces $\Gamma$ to vanish in this limit. More precisely, we define an effective ``sound speed'' $c_\Gamma$ that specifies the scale below which the dark energy fluid is smooth. On super-horizon scales, energy conservation requires $\Gamma$ to satisfy the equation
\begin{equation}
\dot \Gamma = \mathcal{H} \left( S-\Gamma \right) \qquad \frac{c_\Gamma k}{\mathcal{H}} \ll 1 \, ,
\end{equation} 
where $S$ is defined as
\begin{equation}
S \equiv \frac{4 \pi Ga^2}{\mathcal{H}} \left(\rho_{\mathrm{DE}} + p_{\mathrm{DE}} \right) \frac{\theta_{\mathrm{t}}^{\mathrm{N}}}{k^2} \,.
\end{equation}
In the opposite limit, the differential equation must drive $\Gamma$ to zero, i.e.\ we should have $\dot \Gamma \propto -\Gamma$. Interpolating between these two limits provides the differential equation for $\Gamma$,
\begin{equation}
\dot \Gamma = \adotovera \left[S\biggl(1+\frac{c_\Gamma^2 k^2}{\mathcal{H}^2}\biggr)^{-1} - \Gamma \biggl(1+\frac{c_\Gamma^2 k^2}{\mathcal{H}^2}\biggr)\right] \,. \label{eq:Gammadot}
\end{equation}
We can now control the transition scale by choosing $c_\Gamma$, and in practice the value $c_\Gamma \sim 0.4 c_{\mathrm{s}}$ has been found to mimic quintessence dark energy quite well \cite{Fang:2008sn}. This is the standard setting used in \CLASS{} and \CAMB{}, and we shall use the same value in the present work.
Details regarding the computation of $\Gamma$ can be found in e.g.\ \cite{tramPPF}.
For the implementation in $N$-body simulations we shall again also need the pressure perturbation for the PPF dark energy component.
However, this is somewhat more cumbersome to derive than in the fluid model, and therefore we provide a detailed description in Appendix~\ref{appendixA} and \ref{appendixB}.

\subsubsection*{Numerical considerations}
Note that in the limit where $c_\Gamma^2 k^2/\mathcal{H}^2 \gg 1$, \eqref{eq:Gammadot} becomes extremely stiff and therefore numerically challenging for any explicit integrator to solve. However, we also note that in this limit $\Gamma$ is driven to zero and therefore we can simply enforce the condition $\dot{\Gamma}=\Gamma=0$ for values of $c_\Gamma^2 k^2/\mathcal{H}^2$ above some threshold. In \CAMB{} this threshold is set to 30. However, we have found that this threshold may be safely increased well beyond this value, allowing for slightly increased accuracy of the solution without affecting the numerical stability or the computation time in any noticeable way. In particular, we set $\dot{\Gamma}=\Gamma=0$ for $c_\Gamma^2 k^2/\mathcal{H}^2 > 10^4$, while additionally scaling $\dot{\Gamma}$ and $\Gamma$ with a smoothly varying factor between $1$ and $0$ for $10^3 \leq c_\Gamma^2 k^2/\mathcal{H}^2 \leq 10^4$, as to not introduce discontinuities into the system.

In order to obtain precise values for $\delta\rho_\gamma$ and $\delta\rho_\nu$ as needed for \eqref{eq:Poissson_GR} from \CLASS{}, $l_{\text{max}}$ for the photons and neutrinos has to be increased well beyond their standard values. This increases the number of simultaneous differential equations to solve, and so \CLASS{} has to be run using the explicit Runge-Kutta integrator. Thus, introducing the $\dot{\Gamma}=\Gamma=0$ cut-off is vital to keeping the number of time steps required by the code at an acceptable level.

\subsubsection{Dark energy pressure perturbations}
\begin{figure}[t]
\begin{center}
\includegraphics[width=0.8\textwidth]{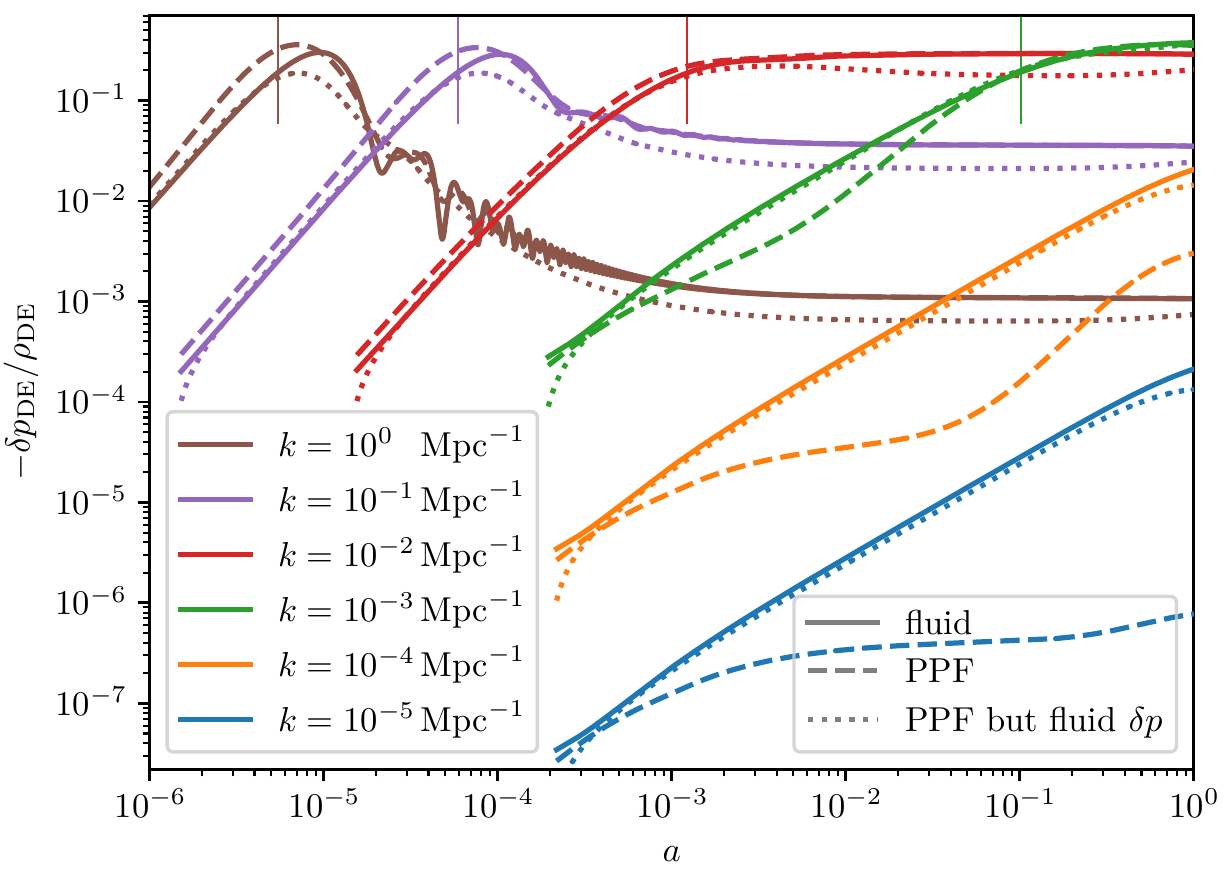}
\end{center}
\caption{\CLASS{} transfer functions for the dark energy pressure perturbation $\delta p_{\text{DE}}$ (in units of the dark energy background density $\rho_{\text{DE}}$) as function of $a$ for different $k$, for the model $w_0 = -0.7$, $w_a = 0$, $c_{\mathrm{s}}^2=1$. Both dark energy as a fluid (solid lines) and using the PPF formalism (dashed lines) are shown. Additionally, the approximate pressure perturbations obtained by plugging the PPF $\delta_{\text{DE}}$ and $\theta_{\text{DE}}$ into the fluid equation \eqref{eq:fluidp} (dotted lines) are shown as well. The thin vertical lines show where $c_\Gamma^2 k^2/\mathcal{H}^2=1$ for each $k$ mode. The transfer functions are given in synchronous gauge.
\label{fig:deltap}}
\end{figure}

In Fig.~\ref{fig:deltap} we show the dark energy pressure perturbation, $\delta p_{\text{DE}}$, for a model with $w_0 = -0.7$, $w_a = 0$, $c_{\mathrm{s}}^2 = 1$ for both the fluid and the PPF parameterisations.
The thin vertical lines show where the term $c_{\Gamma}^2 k^2/\mathcal{H}^2=1$, i.e.\ about where the solution starts to become damped for both the fluid and the PPF solutions. At late times these damped fluid and PPF solutions are very close to identical, though for the largest scales shown this is not the case yet at $a=1$.

As a third option, Fig.~\ref{fig:deltap} also shows the case where the PPF equations are solved to find $\Gamma$ and thus $\delta \rho_{\text{DE}}$, but the pressure perturbation $\delta p_{\text{DE}}$ is calculated using the fluid prescription from \eqref{eq:fluidp}. As can be seen, using this prescription successfully yields good approximate values for the fluid $\delta p_{\text{DE}}$ (at least at early times) even though the dark energy differential equations solved are those of PPF. At later times, the computed $\delta p_{\text{DE}}$ is typically off by a factor of $\sim 2$ compared to the real fluid solution.

On scales beyond the effective sound horizon of the dark energy component, differences between the fluid and PPF pressure perturbation are much larger. For the cases $k = 10^{-5}\,\text{Mpc}^{-1}$ and $k=10^{-4}\,\text{Mpc}^{-1}$ the fluid and PPF prescriptions can yield pressure perturbations which are different by orders of magnitude. Though \eqref{eq:fluidp} can then somewhat successfully be used to obtain the fluid pressure perturbation from the solved PPF system, this is very far from the actual (PPF) pressure perturbation needed. We note that swapping out the much more involved PPF calculations of $\delta p_{\mathrm{DE}}$ presented in Appendices~\ref{appendixA} and \ref{appendixB} for the fluid $\delta p_{\mathrm{DE}}$ \eqref{eq:fluidp} is inconsistent with the Einstein equations and that this inconsistency will show up prominently in the general relativistic correction potential $\gamma$, which in turn leads to large errors in the matter power spectrum on large scales.

\begin{figure}[t]
\begin{center}
\includegraphics[width=0.8\textwidth]{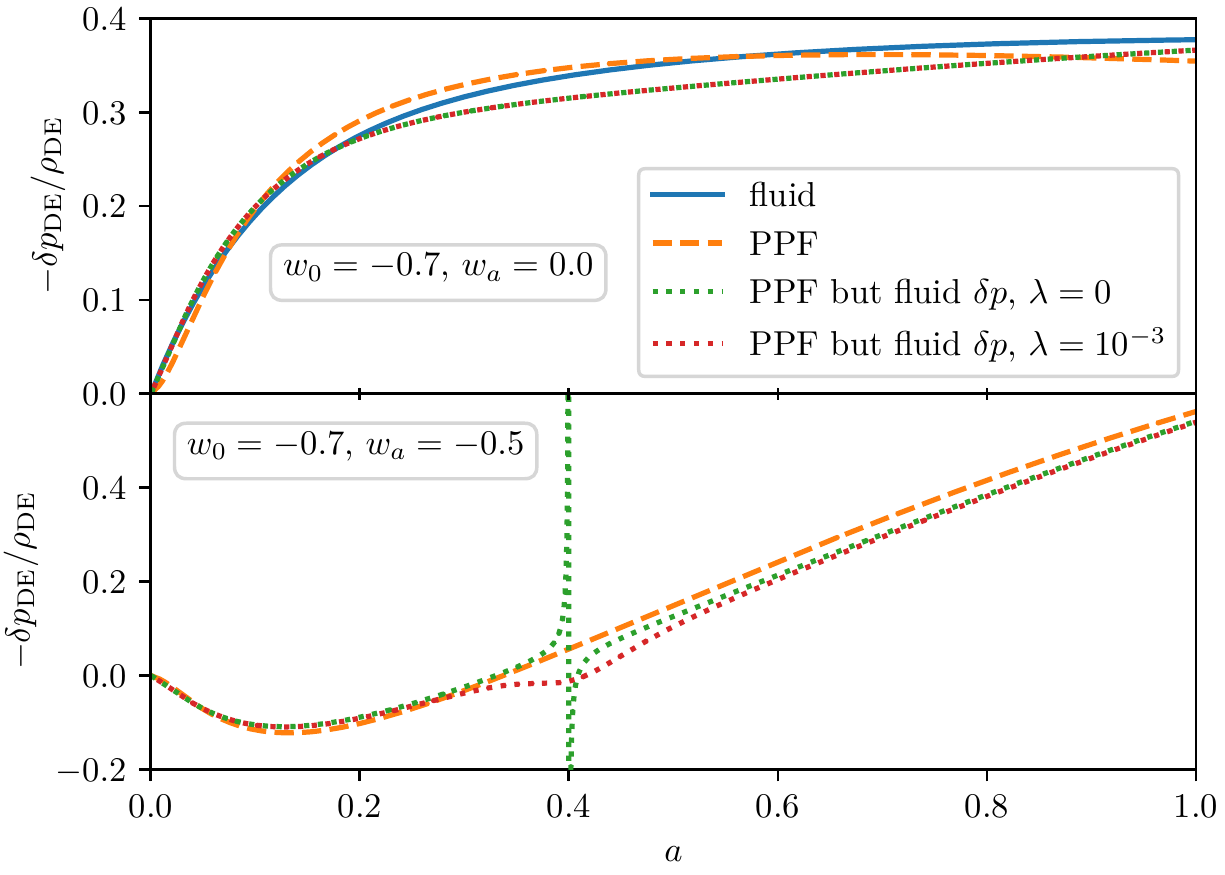}
\end{center}
\caption{\CLASS{} transfer functions for the dark energy pressure perturbation $\delta p_{\text{DE}}$ (in units of the dark energy background density $\rho_{\text{DE}}$) as function of $a$ at $k=10^{-3}\,\mathrm{Mpc}^{-1}$ for two models; one without phantom crossing (top) and one with phantom crossing (bottom). Both panels show the PPF $\delta p_{\text{DE}}$ (dashed lines), while only the top panel shows the fluid $\delta p_{\text{DE}}$ (solid line), as the fluid dark energy equations cannot be integrated across the phantom crossing. Additionally, the approximate pressure perturbations obtained by plugging the PPF $\delta_{\text{DE}}$ and $\theta_{\text{DE}}$ into the fluid equation \eqref{eq:fluidp} with $c_\text{a}^2$ regularised using \eqref{eq:ca2_regularized} (dotted lines) are shown as well. The transfer functions are given in synchronous gauge.
\label{fig:deltap2}}
\end{figure}

In Fig.~\ref{fig:deltap2} we again show the time evolution of the dark energy pressure perturbation. However, in this case we also include a model with phantom crossing (lower panel). As can be seen, the full PPF expression for $\delta p_{\mathrm{DE}}$ shows no divergence at the phantom crossing point at $a=0.4$, while $\delta p_{\mathrm{DE}}$ from \eqref{eq:fluidp} shows the expected pathology. One might consider regularizing the phantom crossing in the fluid case by introducing a small dimensionless parameter, $\lambda$, such that (see (59) of \cite{Kunz:2006wc})
\begin{equation}
c_{\mathrm{a}}^2 = w - \frac{\dot w (1+w)}{3 {\cal H} \left[(1+w)^2 + \lambda \right]}\,, \label{eq:ca2_regularized}
\end{equation}
where $\lambda = 0$ reduces \eqref{eq:ca2_regularized} to the usual $c_{\mathrm{a}}^2 = \dot{p}_{\mathrm{DE}}/\dot{\rho}_{\mathrm{DE}}$. As seen from Fig.~\ref{fig:deltap2}, taking $\lambda > 0$ does indeed regularize the fluid $\delta p_{\mathrm{DE}}$ of \eqref{eq:fluidp}. Though the regularization \eqref{eq:ca2_regularized} leads to a well behaved fluid $\delta p_{\mathrm{DE}}$ even during phantom crossing, this should not be used instead of the properly calculated PPF $\delta p_{\mathrm{DE}}$, as already noted\footnote{We note that the regularization \eqref{eq:ca2_regularized} of \cite{Kunz:2006wc} was not proposed in connection with the PPF parameterisation.}.

\section{Numerical setup and results}

In order to test the formalism outlined above we perform a range of $N$-body simulations using the publicly available \CONCEPT{} $N$-body solver \cite{Dakin:2017idt}. All \CONCEPT{} simulations in this work use cosmological parameters as listed in table~\ref{table:class_parameters}. We use a neutrino sector of three massless neutrinos. The \CONCEPT{} simulations all begin at $a=0.01$, use $1024^3$ matter particles and the potential grids (both $\phi_{\text{sim}}$ and $\phi_{\text{GR}}$) are of size $1024^3$. All \CONCEPT{} simulations are carried out in box sizes of $(65536\,\text{Mpc}/h)^3$, $(8192\,\text{Mpc}/h)^3$ and $(1024\,\text{Mpc}/h)^3$, the power spectra from which are patched together to give the ones presented.

\begin{figure}[t]
\begin{center}
\includegraphics[width=0.95\textwidth]{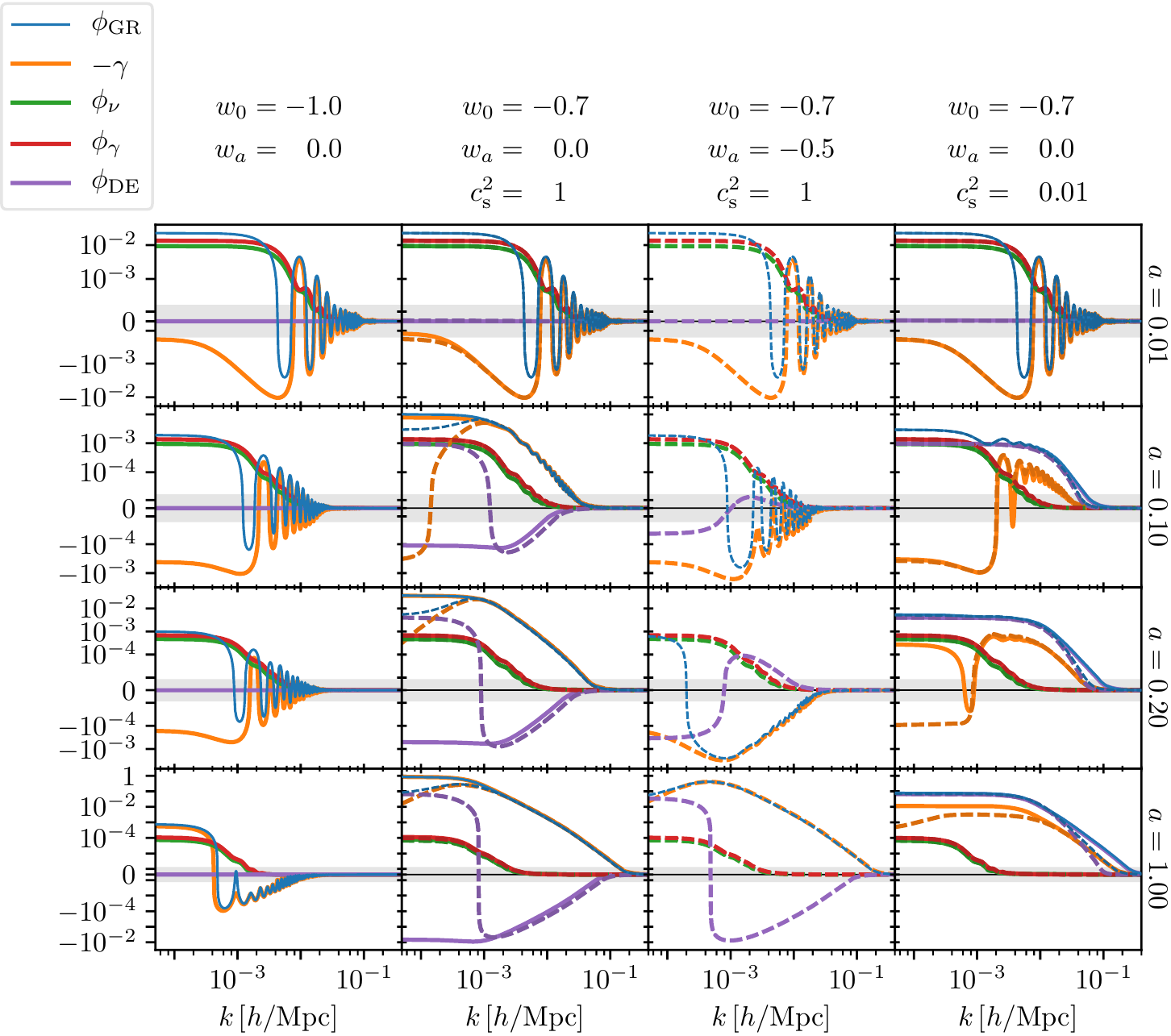}
\end{center}
\caption{Individual contributions to the potential $\phi_{\rm GR}\equiv \phi_\gamma + \phi_\nu + \phi_{\mathrm{DE}} -\gamma^{\text{Nb}}$ at different scale factors, all in $N$-body gauge. The leftmost plots show the case of a cosmological constant, whereas all other plots show the case of $w_0=-0.7$ and different $w_a$ and $c_{\mathrm{s}}^2$. Solid lines result from having DE as a fluid, whereas dashed lines results from the PPF formalism. The grey bands indicate regions where the vertical axes scale linearly.
\label{fig:potential_contributions}}
\end{figure}

\begin{figure}[t]
\begin{center}
\includegraphics[width=0.9\textwidth]{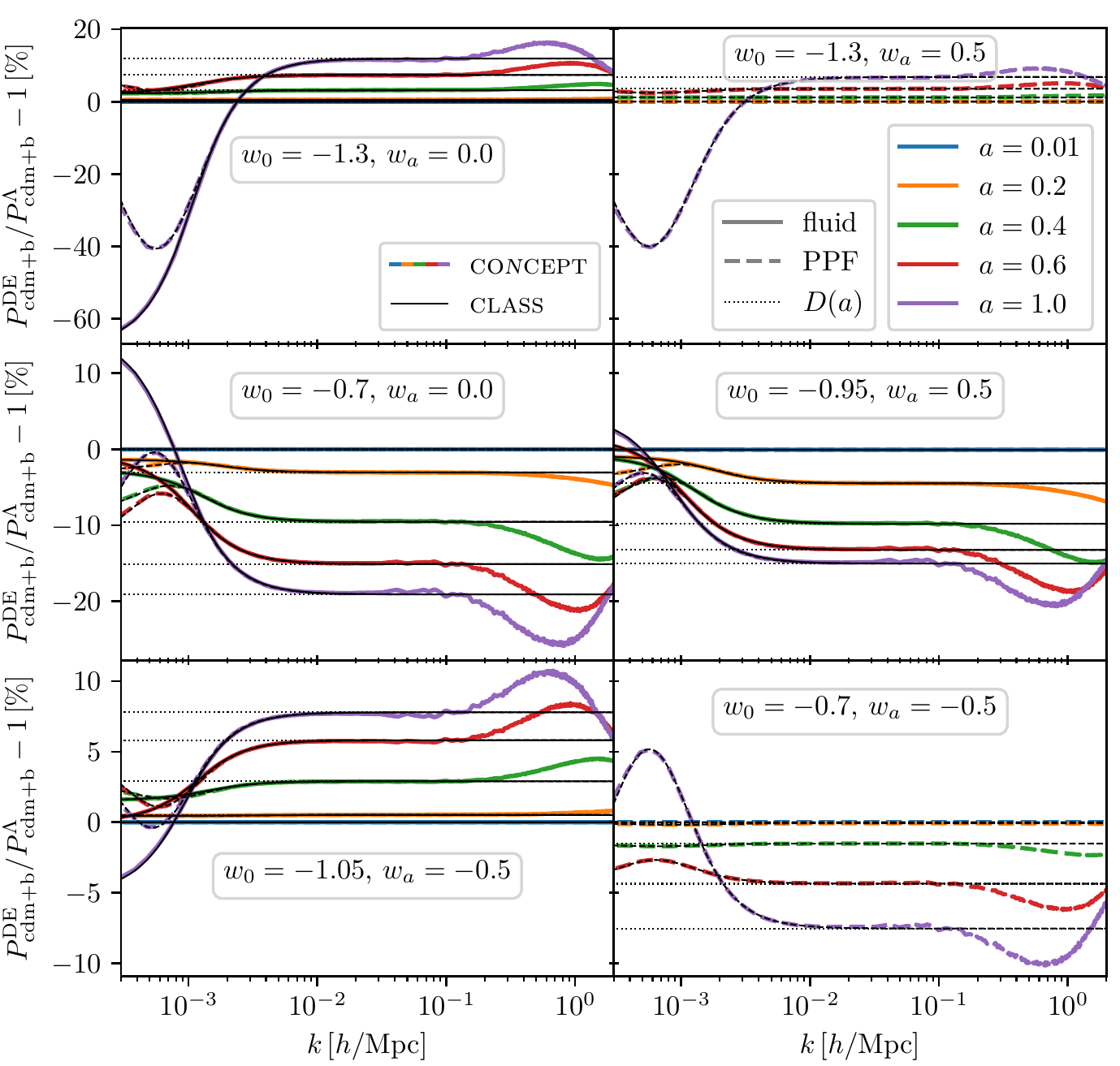}
\end{center}
\caption{Relative matter (cold dark matter and baryons) power spectra between models with dynamical dark energy and a model with a cosmological constant. Coloured lines show results from \CONCEPT{} simulations, whereas black lines indicate the corresponding linear results from \CLASS{}. All plots show dark energy modelled using PPF (dashed lines). Most plots also show dark energy modelled as a fluid (solid lines), the exceptions being those with phantom crossing in the past. All dynamical dark energy models have $c_{\mathrm{s}}^2=1$ and all power spectra are in $N$-body gauge. Finally, the horizontal dotted lines show the relative power as predicted by the linear growth factor $D(a)$.
\label{fig:relative}}
\end{figure}

\begin{figure}[t]
\begin{center}
\includegraphics[width=0.9\textwidth]{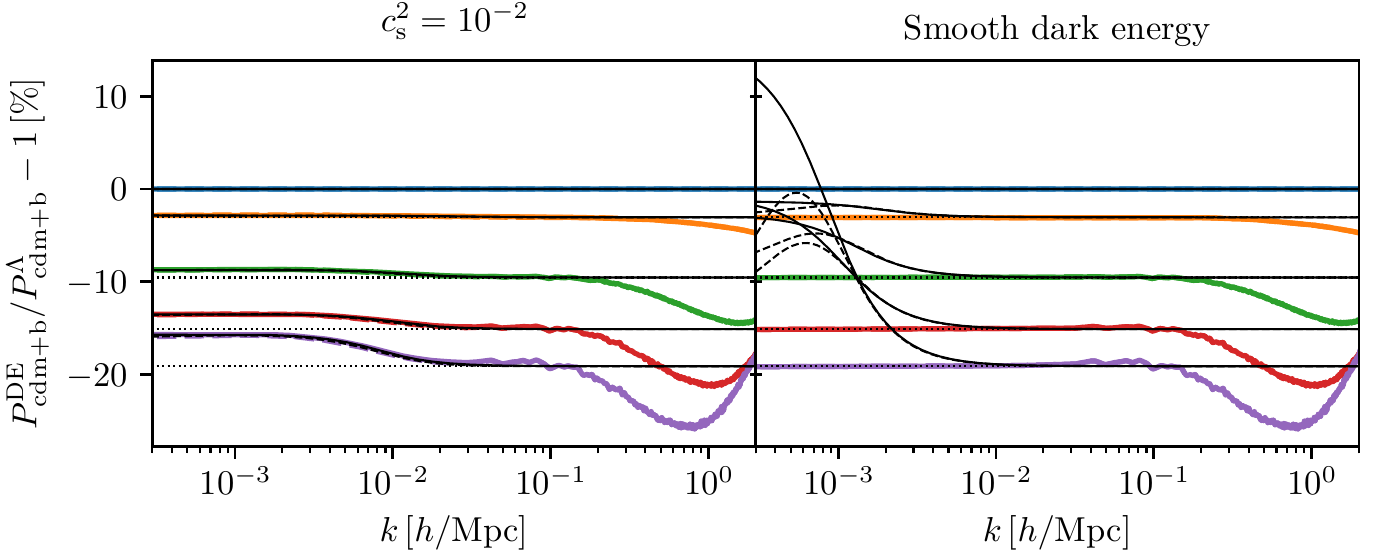}
\end{center}
\caption{Relative matter (cold dark matter and baryons) power spectra between models with $\{w_0=-0.7,\, w_a=0\}$ and a model with a cosmological constant, in $N$-body gauge. All legends from Fig.~\ref{fig:relative} apply. The left plot shows the case of $c_{\mathrm{s}}^2=10^{-2}$. The right plot shows the standard case of $c_{\mathrm{s}}^2=1$, but with the dark energy perturbations left out of the \CONCEPT{} simulations, leading to clear disagreement with linear theory on large scales. Both plots should be compared to the $\{w_0=-0.7,\, w_a=0\}$ panel of Fig.~\ref{fig:relative}.
\label{fig:relative_specials}}
\end{figure}

\begin{table}[tb]
    \begin{center} 
        \begin{tabular}{l c} 
            \hline
            Parameter & Value  \\
            \hline
            $A_\text{s}$  & $2.1 \times 10^{-9}$ \\
            $n_\text{s}$ & $0.96$ \\
            $\tau_\text{reio}$ & $0.0925$  \\
            $\Omega_{\text{b}}$ & $0.049$  \\
            $\Omega_{\text{cdm}}$ & $0.27$ \\
            $h$ & $0.67$ \\
            \hline						
        \end{tabular}
    \end{center}
    \caption{Non-dark energy cosmological parameters for the \CLASS{} runs used.}
    \label{table:class_parameters} 
\end{table}

\begin{table}[tb]
    \begin{center} 
        \begin{tabular}{l c c c c} 
            \hline
            Simulation & $w_0$ & $w_a$ & $c_{\mathrm{s}}^2$ & note \\
            \hline
            A    &    $-1.0\phantom{0}$    &    $\,\,\,\;0.0$ & & (cosmological constant)  \\
            B    &    $-1.3\phantom{0}$    &    $\,\,\,\;0.0$    &    $\phantom{0}1^{\phantom{-2}}$  \\
            C    &    $-1.3\phantom{0}$    &    $\,\,\,\;0.5$    &    $\phantom{0}1^{\phantom{-2}}$  \\
            D    &    $-1.05\phantom{}$    &    $       -0.5$    &    $\phantom{0}1^{\phantom{-2}}$  \\
            E    &    $-0.95\phantom{}$    &    $\,\,\,\;0.5$    &    $\phantom{0}1^{\phantom{-2}}$  \\
            F    &    $-0.7\phantom{0}$    &    $\,\,\,\;0.0$    &    $\phantom{0}1^{\phantom{-2}}$  \\
            G    &    $-0.7\phantom{0}$    &    $       -0.5$    &    $\phantom{0}1^{\phantom{-2}}$  \\
            H    &    $-0.7\phantom{0}$    &    $\,\,\,\;0.0$    &    $10^{-2}$                      \\
            I    &    $-0.7\phantom{0}$    &    $\,\,\,\;0.0$    &    $\phantom{0}1^{\phantom{-2}}$    &    (smooth dark energy) \\
            \hline						
        \end{tabular}
    \end{center}
    \caption{Dark energy cosmological parameters for the simulations.}
    \label{table:class_parametersDE} 
\end{table}

\subsection{Main results}
In Fig.~\ref{fig:potential_contributions} we show the various individual contributions to $\phi_{\text{GR}}$. As expected, we see that the relativistic species (neutrinos and photons) provide the largest contribution to the potential at early times. Later, when dark energy becomes important, the DE perturbations and the DE contribution to the metric potential $\gamma$ dominate completely. This behaviour is seen in all models, regardless of the specific values of $w_0$, $w_a$ and $c_{\mathrm{s}}^2$. Though at late times $|\phi_{\text{DE}}|\gg |\phi_\gamma|, |\phi_\nu|$, what really dominates is the dark energy contribution to $\gamma$. Worth noting here is also that dark energy in the fluid and PPF descriptions, while almost identical for moderate and large $k$, can exhibit very different behaviour for small values of $k$. In particular, the potential contribution $\phi_{\text{DE}}$ for the PPF dark energy tend to switch sign at a given $k$, as seen around $k=10^{-3}\,h/\text{Mpc}$ in the two middle columns of Fig.~\ref{fig:potential_contributions}. This sign change is accompanied with a simultaneous change in $\gamma$, which nullifies and even reverses the overall effect.

In Fig.~\ref{fig:relative} we show relative matter power spectra between models with time-varying DE and a $\Lambda$CDM reference model, corresponding to simulation B--G and A in table~\ref{table:class_parametersDE}, respectively. For reference we have also included the prediction calculated from the difference in the Newtonian growth factor $D(a)$, which comes solely from the change in the background expansion rate. As can be seen from the figure, essentially all models follow the Newtonian linear theory prediction for $k \gtrsim 10^{-2} \, h/{\rm Mpc}$ until the point where they go non-linear at higher $k$-values.
However, surveys with a volume sufficient to probe the region $k \sim 10^{-3}\,\text{Mpc}^{-1}$ will be able to probe differences due to the GR corrections, which can be very large (tens of percent). This will be of particular interest to surveys such as EUCLID and possible future 21-cm surveys with even larger effective volumes \cite{McQuinn:2005hk,Mao:2008ug,Pritchard:2011xb,Sprenger:2018tdb}.

In the left panel of Fig.~\ref{fig:relative_specials} we show a case where the sound speed is smaller than 1, corresponding to simulation H in table~\ref{table:class_parametersDE}. In this case a difference relative to the Newtonian prediction arises around the sound horizon of the dark energy component which is now well inside the current horizon. In the particular case we show, the difference between the Newtonian prediction and the correct result is a few percent already close to $k \sim 10^{-2}\, h/\text{Mpc}$, within the range which can be probed by e.g.\ EUCLID.
Finally, for completeness, the right panel of Fig.~\ref{fig:relative_specials} shows a simulation which has all components \emph{other} than DE correctly implemented, but DE perturbations (including their contributions to $\gamma$) ignored, corresponding to simulation I in table~\ref{table:class_parametersDE}. In that case, we can clearly see that the simulation fails to match the correct linear theory prediction on large scales, but instead continues to follow the prediction of the linear Newtonian growth factor $D(a)$ for all linear scales.

\subsection{Comparison to PKDGRAV3}
\begin{figure}[tb]
\begin{center}
\includegraphics[width=0.9\textwidth]{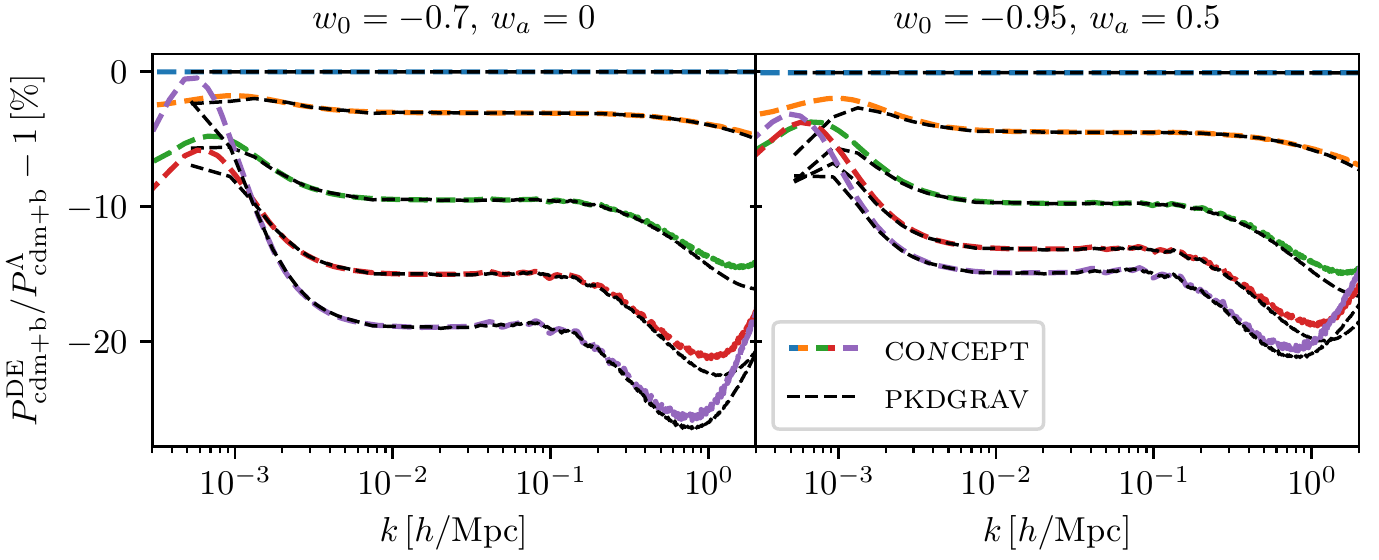}
\end{center}
\caption{Relative matter (cold dark matter and baryons) power spectra between models with PPF dark energy and a model with a cosmological constant. Coloured lines are \CONCEPT{} results and are identical to those shown in the middle row of Fig.~\ref{fig:relative} (consult this figure for the mapping between line colours and values of $a$). Black lines are corresponding relative power spectra produced with \PKDGRAV{}.
\label{fig:pkdgrav}}
\end{figure}

As demonstrated in \cite{Tram:2018znz}, the full framework for adding general relativistic linear theory corrections $\nabla\phi_{\text{GR}}$ to the $N$-body particles of otherwise Newtonian simulations has been successfully implemented into the \PKDGRAV{} code \cite{pkdgrav1,pkdgrav2,pkdgrav3} as well. Fig.~\ref{fig:pkdgrav} shows the same PPF \CONCEPT{} relative matter power spectra as the middle row of Fig.~\ref{fig:relative}, now with the corresponding spectra computed using \PKDGRAV{} added. The agreement is striking for both models, especially considering that \CONCEPT{} and \PKDGRAV{} are very different codes, the first relying on particle-mesh methods and the latter on tree techniques, for computing the Newtonian matter gravity.

At non-linear scales ($k \gtrsim 0.1\, h/\text{Mpc}$), Fig.~\ref{fig:pkdgrav} do show a difference between \CONCEPT{} and \PKDGRAV{}. This has nothing to do with dark energy or the linear corrections in general, as these are completely irrelevant on these scales, as demonstrated by e.g.\ the right panel of Fig.~\ref{fig:relative_specials}. Instead, the difference is produced by the lack of proper small-scale resolution of gravity in \CONCEPT{}.

Now looking to the other end of the spectrum, at the very largest scales ($k \lesssim 10^{-3}\, h/\text{Mpc}$), the \PKDGRAV{} results start to deviate from their \CONCEPT{} (and \CLASS{}, see Fig.~\ref{fig:relative}) counterparts. Here the \PKDGRAV{} simulations fail to obtain convergence of the matter gravity, which is to be expected for a pure tree code\footnote{Due to the very low absolute power at very large scales, the tree has to be crawled extremely deep in order to get these scales correct, essentially converting the $\mathcal{O}(N \log N)$ tree operation into a $\mathcal{O}(N^2)$ direct summation operation.}. As demonstrated by the right panel of Fig.~\ref{fig:relative_specials}, the interesting region regarding dark energy perturbations is $k \lesssim 10^{-2}\,h/\text{Mpc}$, and so the \PKDGRAV{} lines cover about one decade within this region of interest, before convergence is no longer obtainable. During this decade however, exactly the correct behavior is observed.
As an aside then, Fig.~\ref{fig:pkdgrav} in addition demonstrates clearly how pure mesh and pure tree codes have opposite strength/weakness regarding scale resolution.

\section{Discussion}

We have, for the first time, included non-cosmological constant dark energy into $N$-body simulations, not just at the background level, but including perturbations, and fully consistent with GR.

It was shown that these simulations match the solution provided by \CLASS{} exactly in the linear regime, when the potentially highly non-trivial dark energy pressure perturbations are consistently included as a metric source term. 

As expected we find that the dark energy perturbations and their contribution to the metric terms are important on very large scales, and that they become very sub-dominant on smaller scales. We highlight that dark energy in the fluid and PPF descriptions, while almost identical on scales below the dark energy sound horizon, are in general very different on larger scales. In most cases the difference is at scales beyond the reach of currently planned surveys such as EUCLID. However, future 21-cm surveys might be able to reach an effective survey volume large enough to make it detectable (see e.g.\ \cite{McQuinn:2005hk,Mao:2008ug,Pritchard:2011xb,Sprenger:2018tdb}).

We find that for models with $c_{\mathrm{s}}^2=1$ and scales smaller than $k \sim 10^{-2}\, h/\text{Mpc}$ the effect of dark energy is well described solely by its contribution to the background expansion rate. However, for models with smaller dark energy sound speed the dark energy perturbations themselves become important at the percent level already at $k \sim 10^{-2}\, h/\text{Mpc}$, and therefore it might be possible to detect the effect in future surveys (see e.g.\ \cite{Basse:2012wd,Basse:2013zua} for a more detailed discussion).

It is worth noting that even though the effects studied here are mainly relevant on large scales, statistics which correlate short and long wavelengths are susceptible to errors even in the linear regime at small $k$. Examples of this include weak lensing statistics where full lightcones must be properly constructed, as well as statistics used to probe non-Gaussianity in the squeezed limit. 

Finally, we note that even though our current implementation has been done for the fluid and PPF dark energy descriptions, it should work equally well for any dark energy model which couples only gravitationally to other sectors and which contains no non-linear inhomogeneities (e.g.\ quintessence).

\section*{Acknowledgements}
JD, SH, and TT are supported by the Villum Foundation. MK is supported by Swiss National Science Foundation grant 200021\_182748 ``The Euclid Emulator Project''. We would like to thank Doug Potter and Hugues de Laroussilhe for implementing the linear species mesh force in \PKDGRAV{}.

\appendix
\section{Computing $\delta p_\text{DE}$ in the PPF formalism using numerical differentiation}\label{appendixA}
The dark energy pressure perturbation in the PPF formalism can be computed from the dark energy continuity equation. Following \CLASS{}, we define $m_\text{cont.}^{\mathrm{s}} \equiv \dot h/2$ and  $m_\text{cont.}^{\mathrm{N}} \equiv -3\dot \phi$. The continuity equation can then be written uniformly in Newtonian and synchronous gauge as
\begin{equation}
\partial_\tau \delta \rho_\text{DE} = - (\rho_\text{DE} + p_\text{DE}) \left(\theta_\text{DE} + m_\text{cont.} \right) -3 \mathcal{H} \left(\delta \rho_\text{DE} + \delta p_\text{DE} \right)\, ,
\end{equation}
from where it follows that
\begin{equation}
\delta p_\text{DE} = -\delta \rho_\text{DE} - \frac{1}{3 \mathcal{H}} \bigl[ \partial_\tau \delta \rho_\text{DE}  + (\rho_\text{DE} + p_\text{DE})\left(\theta_\text{DE} + m_\text{cont.} \right) \bigr] \,.
\end{equation}
As $\partial_\tau \delta\rho_\text{DE}$ is not solved for by \CLASS{}, this has to be found through numerical differentiation. Though doable, superior accuracy can be obtained by writing $\delta p_\text{DE}$ as a purely algebraic expression in terms of known quantities. For this, see Appendix~\ref{appendixB}.

\section{Computing $\delta p_{\text{DE}}$ in the PPF formalism algebraically}\label{appendixB}
Without relying on numerical differentiation as in Appendix~\ref{appendixA}, the dark energy pressure perturbation in the PPF formalism is highly non-trivial to compute. To ease the notation, we start by defining some quantities:
\begin{align}
x &\equiv \frac{c_\Gamma^2 k^2}{a^2 H^2} \, , &  y&\equiv \frac{9}{2} \frac{a^2}{k^2} \left( \rho_{\mathrm{t}} + p_{\mathrm{t}} \right) \, , & z&\equiv \frac{2}{3}\frac{k^2 H}{a} \,, \\
\dot x &\equiv -2x \left[\adotovera + \frac{\dot H}{H} \right] \, , &  \dot y&\equiv y \left[ 2\adotovera+ \frac{\dot \rho_{\mathrm{t}} + \dot p_{\mathrm{t}}}{\rho_{\mathrm{t}} + p_{\mathrm{t}}} \right] \, , & \dot z&\equiv z \left[ \frac{\dot H}{H} -\adotovera \right] \,,
\end{align}
where $H\equiv\dot{a}/a^2$ is the Hubble parameter.

In terms of $x$, $y$ and $z$ we can write the PPF formulae as
\begin{align}
S &\equiv z^{-1} \left(\rho_{\mathrm{DE}} + p_{\mathrm{DE}} \right) \theta_{\mathrm{t}}^{\text{N}} \, , \\
\dot \Gamma &= \adotovera \left[ \frac{S}{1+x} -\Gamma (1+x) \right] \, , \\
(\rho_{\mathrm{DE}} +p_{\mathrm{DE}}) \theta_{\mathrm{DE}} &= (\rho_{\mathrm{DE}} +p_{\mathrm{DE}}) \theta_{\mathrm{t}} - \frac{z}{1+y}  \left[\frac{S}{1+x^{-1}} + \Gamma x\right] \nonumber \\
 &=(\rho_{\mathrm{DE}} +p_{\mathrm{DE}}) \bigl(\theta_{\mathrm{t}}^{\text{N}}-k^2\alpha \bigr) - \frac{z}{1+y}  \left[\frac{S}{1+x^{-1}} + \Gamma x\right]  \nonumber \\
&= z \left(S-(1+y)^{-1} \left[ \frac{S}{1+x^{-1}} + \Gamma x \right] \right) -k^2 \alpha \left(\rho_{\mathrm{DE}} + p_{\mathrm{DE}} \right)\, , \label{eq:thetaeexpr}
\end{align}
where $\alpha = (\dot h + 6 \dot \eta)/2k^2$ in synchronous gauge and $\alpha=0$ in Newtonian gauge. We can write the Euler equation in both gauges as
\begin{align}
\partial_\tau (\rho_{\text{DE}}+p_{\text{DE}}) \theta_{\text{DE}}  = &-4 \mathcal{H} (\rho_{\text{DE}}+p_{\text{DE}}) \theta_{\text{DE}} + k^2 \left[\delta p_{\text{DE}} - (\rho_{\text{DE}}+p_{\text{DE}}) \sigma_{\text{DE}} \right] \notag \\
&+ (\rho_{\text{DE}}+p_{\text{DE}}) m_\text{Euler} \,, \label{eq:Euler}
\end{align}
where, as in \CLASS{}, $m^{\text{N}}_\text{Euler}=k^2\psi$ and $m^{\text{s}}_\text{Euler}=0$. Since $\sigma_{\mathrm{DE}}\equiv 0$, the PPF pressure perturbation can be written as
\begin{equation}\label{eq:deltapEuler}
\delta p_{\mathrm{DE}} = \frac{1}{k^2} \Bigl( \partial_\tau (\rho_{\mathrm{DE}}+p_{\mathrm{DE}}) \theta_{\mathrm{DE}} + 4\mathcal{H} (\rho_{\mathrm{DE}} + p_{\mathrm{DE}}) \theta_{\mathrm{DE}} -(\rho_{\mathrm{DE}} + p_{\mathrm{DE}}) m_\text{Euler} \Bigr) \, .
\end{equation}
All quantities in \eqref{eq:deltapEuler} are readily available in \CLASS{}, except for the time derivative $\partial_\tau (\rho_{\mathrm{DE}}+p_{\mathrm{DE}}) \theta_{\mathrm{DE}}$ which we shall now construct by taking the derivative of \eqref{eq:thetaeexpr}:
\begin{align}
\partial_\tau (\rho_{\mathrm{DE}} + p_{\mathrm{DE}}) \theta_{\mathrm{DE}} &= \dot z \left(S-(1+y)^{-1} \left[ \frac{S}{1+x^{-1}} + \Gamma x \right] \right)  + z \left( \dot S +\frac{\dot y}{(1+y)^2}  \left[ \frac{S}{1+x^{-1}} + \Gamma x \right]  \right. \nonumber \\
& \phantom{=z\biggl(} \left.-(1+y)^{-1}\left[\frac{\dot S}{1+x^{-1}} +\frac{S \dot x}{(1+x)^2} + \dot \Gamma x + \Gamma \dot x \right]        \right) \nonumber \\
&\phantom{=} -k^2\dot \alpha (\rho_{\mathrm{DE}} + p_{\mathrm{DE}}) - k^2 \alpha (\dot \rho_{\mathrm{DE}} + \dot p_{\mathrm{DE}}) \, . \label{eq:thetaederivative}
\end{align}
To evaluate \eqref{eq:thetaederivative} we must compute $\dot S$: 
\begin{align}
\dot S &= \partial_\tau \left[ z^{-1} (\rho_{\mathrm{DE}} + p_{\mathrm{DE}}) (\theta_{\mathrm{t}}+k^2 \alpha) \right] \nonumber \\
&=-\frac{\dot z}{z} S + z^{-1} (\dot \rho_{\mathrm{DE}} + \dot p_{\mathrm{DE}} )  (\theta_{\mathrm{t}}+k^2 \alpha)  +  z^{-1} (\rho_{\mathrm{DE}} + p_{\mathrm{DE}} )  (\dot \theta_{\mathrm{t}}+k^2 \dot \alpha) \, .
\end{align}
All that remains is to evaluate $\dot \theta_{\mathrm{t}}$ through the Euler equation:
\begin{align}
 \dot \theta_\mathrm{t} &= -\left( \adotovera (1-3w_\mathrm{t}) +\frac{\dot w_\mathrm{t}}{1+w_\mathrm{t}} \right) \theta_\mathrm{t} +\frac{k^2 \delta p_\mathrm{t}}{\rho_\mathrm{t} + p_\mathrm{t}} -k^2 \sigma_\mathrm{t} + m_\text{Euler} \nonumber \\
 &= -\adotovera \theta_\mathrm{t} - \frac{\dot p_\mathrm{t}}{\rho_\mathrm{t} + p_\mathrm{t}}  \theta_\mathrm{t} +\frac{k^2 \delta p_\mathrm{t}}{\rho_\mathrm{t} + p_\mathrm{t}} -k^2 \sigma_\mathrm{t} + m_\text{Euler} \nonumber \\
  &= -\adotovera \theta_\mathrm{t} -\left(\dot p_\mathrm{t} \theta_\mathrm{t} -k^2 \delta p_\mathrm{t} +k^2 (\rho_\mathrm{t} + p_\mathrm{t}) \sigma_\mathrm{t}\right) \frac{1}{\rho_\mathrm{t} + p_\mathrm{t}} + m_\text{Euler} \, .
\end{align}
Note that $\sigma_\mathrm{t}$ is the total anisotropic stress $\sigma_\mathrm{tot}$, since anisotropic stress is absent for the PPF fluid. In synchronous gauge we also need $\dot \alpha$ which we can evaluate from the Einstein equation involving $\sigma$ as
\begin{equation}
\dot \alpha = -2 \adotovera \alpha + \eta -\frac{9}{2} \frac{a^2}{k^2} (\rho_\mathrm{tot} + p_\mathrm{tot})\sigma_\mathrm{tot}\,.
\end{equation}
In Newtonian gauge we can construct $\psi$ from the evolution variable $\phi$ and the total anisotropic stress as
\begin{equation}
\psi = \phi-\frac{9}{2} \frac{a^2}{k^2} (\rho_\mathrm{tot} + p_\mathrm{tot}) \sigma_\mathrm{tot} \,.
\end{equation}

\bibliographystyle{utcaps}

\providecommand{\href}[2]{#2}\begingroup\raggedright
\endgroup

\end{document}